\documentclass[aps,prx,preprint,superscriptaddress,floatfix]{revtex4-2} 
\usepackage{blindtext} 
\usepackage{graphicx}
\usepackage{eqnarray}
\usepackage{amsmath}
\usepackage{xcolor}
\usepackage{appendix}

\begin{document}

\title{Phase-matching in high-order harmonic generation with a plasma-reshaped driving field}

\author{Gaspard Beaufort}
\email[Contact author: ]{gaspard.beaufort@fysik.lu.se}
\affiliation{Department of Physics, Lund University, P.O. Box 118, SE-22100 Lund, Sweden}
\author{Vénus Poulain}
\affiliation{Department of Physics, Lund University, P.O. Box 118, SE-22100 Lund, Sweden}
\author{Melvin Redon}
\affiliation{Department of Physics, Lund University, P.O. Box 118, SE-22100 Lund, Sweden}
\author{Robin Weissenbilder}
\affiliation{Department of Physics, Lund University, P.O. Box 118, SE-22100 Lund, Sweden}
\author{Elisa Appi}
\affiliation{Department of Physics, Lund University, P.O. Box 118, SE-22100 Lund, Sweden}
\author{Filip Salbrink}
\affiliation{Department of Physics, Lund University, P.O. Box 118, SE-22100 Lund, Sweden}
\author{Saga Westerberg}
\affiliation{Department of Physics, Lund University, P.O. Box 118, SE-22100 Lund, Sweden}
\author{Aimee J. Ross}
\affiliation{ASML Research, ASML Netherlands B.V., 5504 DR Veldhoven, Netherlands}
\author{Sylvianne Roscam Abbing}
\affiliation{ASML Research, ASML Netherlands B.V., 5504 DR Veldhoven, Netherlands}
\author{David P. O'Dwyer}
\affiliation{ASML Research, ASML Netherlands B.V., 5504 DR Veldhoven, Netherlands}
\author{Yin Tao}
\affiliation{ASML Research, ASML Netherlands B.V., 5504 DR Veldhoven, Netherlands}
\author{Per~Eng-Johnsson}
\affiliation{Department of Physics, Lund University, P.O. Box 118, SE-22100 Lund, Sweden}
\author{Cord L. Arnold}
\affiliation{Department of Physics, Lund University, P.O. Box 118, SE-22100 Lund, Sweden}
\author{Anne L'Huillier}
\affiliation{Department of Physics, Lund University, P.O. Box 118, SE-22100 Lund, Sweden}
\author{Peter Smorenburg}
\affiliation{Department of Physics, Lund University, P.O. Box 118, SE-22100 Lund, Sweden}
\affiliation{ASML Research, ASML Netherlands B.V., 5504 DR Veldhoven, Netherlands}
\author{Anne-Lise Viotti}
\affiliation{Department of Physics, Lund University, P.O. Box 118, SE-22100 Lund, Sweden}

\date{\today}

\begin{abstract}

The transition from titanium-sapphire to ytterbium-based laser technology enables turnkey implementations of high-order harmonic generation (HHG) setups at high repetition rates, opening up new applications, e.g., in semiconductor metrology. This, however, requires exploring and understanding the impact of new experimental conditions. Here, we study HHG in neon over the spectral range 8-17~nm with high spectral resolution, using a post-compressed ytterbium laser. Our measurements reveal a large pulse-energy- and pressure-dependent spectral shift, spanning more than two harmonic orders. The femto-chirp of the harmonic emission combined with the blueshift of the fundamental field allows us to relate the observed spectral shifts to the temporal window of harmonic generation, which is determined by phase-matching. To understand these dynamics, it is essential to consider the reshaping of the driving laser field inside the plasma, despite the low driving intensity and long pulse duration. We introduce an on-axis phase-matching description that captures these effects and reproduces the experimental and simulated trends. These results demonstrate the central role of plasma-induced laser reshaping in phase-matching of HHG, which should be accounted for when optimizing sources in a multi-parameter landscape.


\end{abstract}

\maketitle

\section{Introduction}

High-order harmonic generation (HHG) in gases \cite{mcphersonStudiesMultiphotonProduction1987, ferrayMultipleharmonic1988} has become a standard approach for producing coherent extreme-ultraviolet (XUV) radiation in table-top setups. The resulting attosecond light pulses are crucial for studying matter on ultrashort timescales \cite{mikaelssonHighrepetitionRateAttosecond2021, kretschmarAttosecondInvestigationExtremeultraviolet2022, haesslerPhaseresolvedAttosecondNearthreshold2009}, while the short wavelength of the emitted light makes it useful for coherent diffractive imaging \cite{zurchRealtimeSubwavelengthUltrafast2014} and nanostructure characterization \cite{jiaOpticalHighorderHarmonic2020}. Over the past decades, extensive work has therefore focused on improving the efficiency \cite{constantOptimizingHighHarmonic1999a,gaardeMacroscopic2008,weissenbilderHowOptimizeHighorder2022} of HHG sources. These efforts have led to a broad understanding of the single‑atom response as well as the macroscopic phase-matching mechanisms that, together, shape the XUV emission.

Despite this progress, optimizing HHG remains challenging because the process depends on many different parameters, such as laser intensity, wavelength, pulse duration, focusing geometry, atomic species, gas pressure, and medium length \cite{constantOptimizingHighHarmonic1999a,shinerWavelength2009, weissenbilderHowOptimizeHighorder2022, westerbergInfluenceLaserPulse2025,klasUnleashingHHGEfficiency2026,popmintchevBright2012}. On the macroscopic scale, efficient HHG requires constructive interference of the XUV radiation generated throughout the medium. Phase-matching is achieved if the XUV field emitted from the driving laser field remains in phase with the XUV field already generated and propagating through the medium. This requires that the positive contribution to the phase-mismatch from neutral-atom dispersion and the negative contributions from plasma dispersion and focusing geometry cancel each other \cite{weissenbilderHowOptimizeHighorder2022}, which is not always possible, especially at high photon energy \cite{popmintchevBright2012, klasUnleashingHHGEfficiency2026}. In the conventional description, the driving laser is assumed to remain close to a Gaussian beam and pulse during propagation. However, efficient HHG can also be achieved if the driving field is reshaped in space and time as it propagates through the plasma. This has been demonstrated at peak intensities above $10^{15}$~W/cm$^2$, where the medium can become strongly ionized \cite{sunExtended2017, johnsonHighfluxSoftXray2018, rivasPropagationenhanced2018, majorHighorderHarmonicGeneration2023, azzolinEngineeringHighorderHarmonic2026, ansariBroadband2026a}, as well as for few-cycle driving pulses, where the rapid temporal evolution of the plasma strongly reshapes the field during the pulse \cite{majorEffect2019, schotzPhaseMatching2020, fuExtension2022, chenPhaseMatchedHighHarmonicGeneration2023}. In these conditions, the spatio-temporal reshaping of the fundamental field modifies the phase-matching dynamics and can enable efficient HHG with high photon energy or the generation of isolated attosecond pulses \cite{laiInfluence2011, schotzPhaseMatching2020, qinEnhanced2023, chienFilamentationassistedIsolatedAttosecond2026, tangOptimal2021}. 

In this work, we investigate HHG in a neon gas jet, using a post-compressed 45~fs-long ytterbium laser at a central wavelength of 1032~nm. The peak intensity of the laser in vacuum stays below $7\times10^{14}$~W/cm$^2$, the gas jet is 600~$\mu$m long (0.2 times the Rayleigh length) with a pressure below 500~mbar, so that the estimated degree of ionization in the medium stays typically below 5~\%. We record high-resolution harmonic spectra up to the 119$^{\text{th}}$ order as a function of the driving-pulse energy and the gas-jet pressure. We observe spectral shifts spanning more than two harmonic orders, and ultra-broadband spectra for certain combinations of pulse energy and jet pressure. These spectral shifts can be explained by two time-dependent mechanisms. The first is the dipole phase, the phase accumulated by the electron wave packet during its propagation in the continuum  \cite{gaardeSpatiotemporalSeparationHigh1999,gaardeQuantumPathDistributions2002}. Because the dipole phase depends on the laser intensity, which varies throughout the pulse, the emitted harmonics acquire a time-dependent instantaneous frequency, which is known as the femto-chirp \cite{varju||FrequencyChirpHarmonic2005}. The second mechanism is the plasma-induced blueshift of the driving laser, which results from the increase of the free-electron density during the pulse and the associated change in the refractive index of the medium \cite{altucciTunable1999, khokhlovaRole2024}. Together, these two mechanisms establish a time-to-frequency mapping, where the spectral widths and shifts of the harmonics can be interpreted in the time domain, unraveling the dynamics of phase-matching in the nonlinear medium. 

Despite the multi-cycle driving pulses with moderate intensities, and the short gas jet with modest pressure considered here, the interpretation of our experimental results requires accounting for the plasma-induced reshaping of the fundamental field, which dominates the phase-matching dynamics. We successfully compare the measurements with propagation simulations and propose an on-axis description of the different phase-mismatch contributions that incorporates the reshaping of the driving field. This framework provides an intuitive picture of how reshaping modifies the different contributions to phase-matching. 

In Section \ref{sec:Theoretical_Framework}, we first introduce the relevant theoretical tools used in this work, explaining the time-to-frequency mapping and describing on-axis phase-matching in the Gaussian and reshaped cases. We then present the experimental and numerical methods and results in Section~\ref{sec:Methods}. Finally, Section~\ref{sec:Interpretation} describes the interpretation of the observed spectra.

\section{\label{sec:Theoretical_Framework}Theoretical framework}

HHG results from an interplay between the microscopic response of individual atoms to the driving laser and macroscopic propagation effects within the generating medium. In this section, we develop the theoretical framework used throughout the paper. We first describe how the temporal window of the harmonics is encoded in the observed harmonic spectrum through the combined effects of the dipole phase induced femto-chirp and the plasma-induced blueshift of the driving field. We then discuss phase-matching in the case of a Gaussian and a plasma-reshaped fundamental beam, which imposes a temporal window for the harmonic generation.

\subsection{Dipole phase}

The three-step model gives an intuitive and simple description of the single atom response to the driving laser \cite{lewensteinTheory1994}. An electron first tunnels out of the atom, is accelerated in the continuum by the laser field and may recombine to the parent ion,  emitting the acquired kinetic energy as an XUV photon. For harmonics on the plateau, the dominant contribution to the dipole response is associated with the two shortest trajectories that lead to the same kinetic energy upon recombination, commonly referred to as the ``short" and ``long" trajectories.

During this process, the electron wave-packet acquires a phase, which is called the dipole phase, that is imprinted on the emitted XUV field. For the harmonics in the plateau region, a linear approximation of the time-to-energy curve can be interpreted as a group delay and integrated to get the expression of the phase as in ref.~\cite{guoPhaseControlAttosecond2018}. For the short trajectory, the intensity-dependent phase term can be approximated as follows:
\begin{equation}
    \phi_{\mathrm{dip}}(\Omega) \approx \frac{\gamma}{I}(\Omega-\Omega_p)^2.
    \label{eq:phi_dip}
\end{equation}
Here, $I$ is the laser intensity, $\Omega=q\omega_0$ is the frequency of the harmonic $q$ with $\omega_0$ the frequency of the fundamental laser, $\Omega_p=I_p/\hbar$ is the ionization energy frequency with $I_p$ the ionization potential of the atom. The constant $\gamma$ is $0.22 c m_e/(\alpha \lambda_0)$, where $c$ is the speed of light in vacuum, $m_e$ the electron mass, $\alpha$ the fine structure constant and $\lambda_0$ the central wavelength of the laser \cite{weissenbilderHowOptimizeHighorder2022}. This approximation is more general than the commonly used $\phi_\mathrm{dip}(\Omega)=\alpha (\Omega) I$, which is only valid over a limited intensity range and for a given frequency $\Omega$ \cite{lewensteinPhase1995, schotzPhaseMatching2020}.

\subsection{Time-to-frequency mapping in HHG}

In the following, we consider the temporal profile of an individual harmonic order, obtained as the inverse Fourier transform of its spectrum. This temporal profile should not be confused with the train of attosecond pulses corresponding to the full harmonic spectrum. The instantaneous frequency of the harmonic evolves throughout the driving pulse, establishing a time-to-frequency mapping.

\begin{figure}[htbp]
    \centering
    \includegraphics[width=0.99\linewidth]{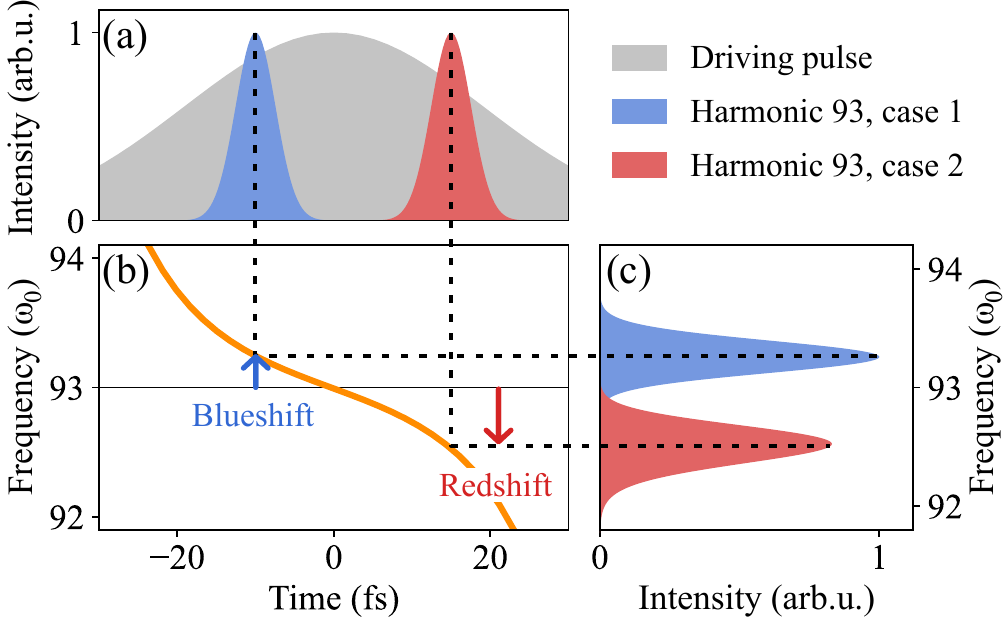}
    \caption{Effect of the femto-chirp of the short trajectory on the harmonic spectrum. (a) Temporal profile of harmonic 93 in two different emission windows together with the driving laser pulse. The driving intensity is $5\times10^{14}$~W/cm$^2$ at a pulse duration of 45~fs. (b) Instantaneous frequency of harmonic 93, considering only the femto-chirp. (c) Spectrum of harmonic 93 in the two different cases. A harmonic emitted during an increase of the laser intensity is blueshifted (case 1), while a harmonic emitted during a decrease of the laser intensity is redshifted (case 2).}
    \label{fig:tf_mapping}
\end{figure}

Two mechanisms contribute to this time-to-frequency mapping. The first is the intensity dependence of the dipole phase [Eq.~(\ref{eq:phi_dip})], which gives rise to the so-called femto-chirp \cite{varju||FrequencyChirpHarmonic2005, salieresCoherence1995, changTemporal1998}. The phase accumulated by the emitted XUV field changes as the intensity evolves throughout the laser pulse, shifting the instantaneous frequency of the harmonic over time. The resulting spectral shift of the $q$-th harmonic is given by
\begin{equation}
    \Delta\Omega_{\mathrm{dip}}(t) = -\frac{\partial \phi_{\mathrm{dip}}}{\partial t} \approx -\frac{\partial \phi_{\mathrm{dip}}}{\partial I}\frac{\partial I}{\partial t}.
\end{equation}
For both short and long trajectories, the dipole phase [Eq.~(\ref{eq:phi_dip})] decreases with increasing intensity (${\partial \phi_{\mathrm{dip}}/\partial I<0}$), so that the sign of the frequency shift is determined by the intensity gradient $\partial I/\partial t$. Harmonics emitted on the rising edge of the pulse are blueshifted, whereas they are redshifted on the falling edge \cite{picotAsymmetricPolarizationGating2025, heylSpectrallyResolvedMaker2011,ouahiouneSubcycle2026}, as illustrated in Fig.~\ref{fig:tf_mapping}.

The second contribution originates from the propagation of the driving laser pulse through the ionizing gas. As free-electrons accumulate during the pulse, the refractive index quickly decreases, producing an accumulated blueshift of the fundamental field $\Delta\omega_0(r,z,t)$. This blueshift is strongest at the peak of the pulse, where the ionization rate is the highest, and increases along the propagation direction as the laser travels through the plasma \cite{khokhlovaRole2024}. This shift is then transferred to the $q$-th harmonic at the spatio-temporal point $(r,z,t)$ of generation.

The instantaneous frequency of a given harmonic $q$ can therefore be expressed as \cite{nefedovaDeterminationSpectralVariation2018}:
\begin{equation}
    \Omega_{\text{inst}}(r,z,t) = q\omega_0 + q\Delta\omega_0(r,z,t)+\Delta\Omega_{\text{dip}}(r,z,t).
    \label{eq:Om_inst}
\end{equation}
The harmonic field is also directly affected by the propagation in the plasma, but the corresponding blueshift is negligible compared to the other contributions. Equation~(\ref{eq:Om_inst}) therefore establishes a time-to-frequency mapping. As discussed in the next section, phase-matching selects the temporal window over which harmonic emission is efficient. Consequently, the measured harmonic spectrum is a spectral signature of this selected temporal window.


\subsection{\label{sec:PM_Gaussian}Phase-matching of harmonics with a Gaussian field}

The HHG efficiency does not only depend on the response of a single atom but also on how the XUV emission builds up during propagation. The XUV intensity is maximized when the phase of the emitted field matches the phase accumulated by the already propagating field over the interaction length. This condition is achieved when the phase-mismatch vector $\Delta k=qk_1-k_q$ is sufficiently small, where $k_1$ is the wave-vector of the fundamental electric field and $k_q$ the one of the harmonic $q$. The total phase-mismatch includes contributions from dispersion due to the neutral atoms $\Delta k_{\text{at}}>0$, dispersion due to the free electrons in the plasma $\Delta k_{\text{fe}}<0$, the geometric phase of the driving laser beam $\Delta k_{\text{geom}}$ and the dipole phase-mismatch $\Delta k_{\mathrm{dip}}$. The sum of these terms has to cancel out in order to achieve phase-matching:
\begin{equation}
   \Delta k_{\text{at}} + \Delta k_{\text{fe}} + \Delta k_{\text{geom}} + \Delta k_{\mathrm{dip}} =  0.
   \label{eq:delta k=0}
\end{equation}

For a Gaussian beam without reshaping, the plasma-dispersion term $\Delta k_{\text{fe}}$ and the geometrical contribution associated with the Gouy phase $\Delta k_{\text{geom}}$ are both negative. Efficient phase-matching therefore requires positive contributions from neutral-atom dispersion $\Delta k_{\text{at}}$, or from the dipole phase-mismatch $\Delta k_{\text{dip}}$ when the gas jet is placed after the focus. As the degree of ionization $\eta_{\mathrm{fe}}(t)$ increases during the laser pulse, the plasma-dispersion term $\Delta k_{\text{fe}}(t)\propto-\eta_{\mathrm{fe}}(t)$ becomes increasingly negative, so that phase-matching is only achieved during a limited temporal window, in which the different contributions compensate for each other.

Similarly, increasing the driving intensity enhances ionization, making the plasma-dispersion term more negative. Above a certain critical ionization, this contribution cannot be compensated for by the remaining positive terms anymore, and phase-matching becomes impossible \cite{popmintchevBright2012}. Thus, conventional HHG is expected to occur only within a limited intensity window.

\subsection{Plasma-induced spatial reshaping}

\begin{figure}[htbp]
    \centering
    \includegraphics[width=0.99\linewidth]{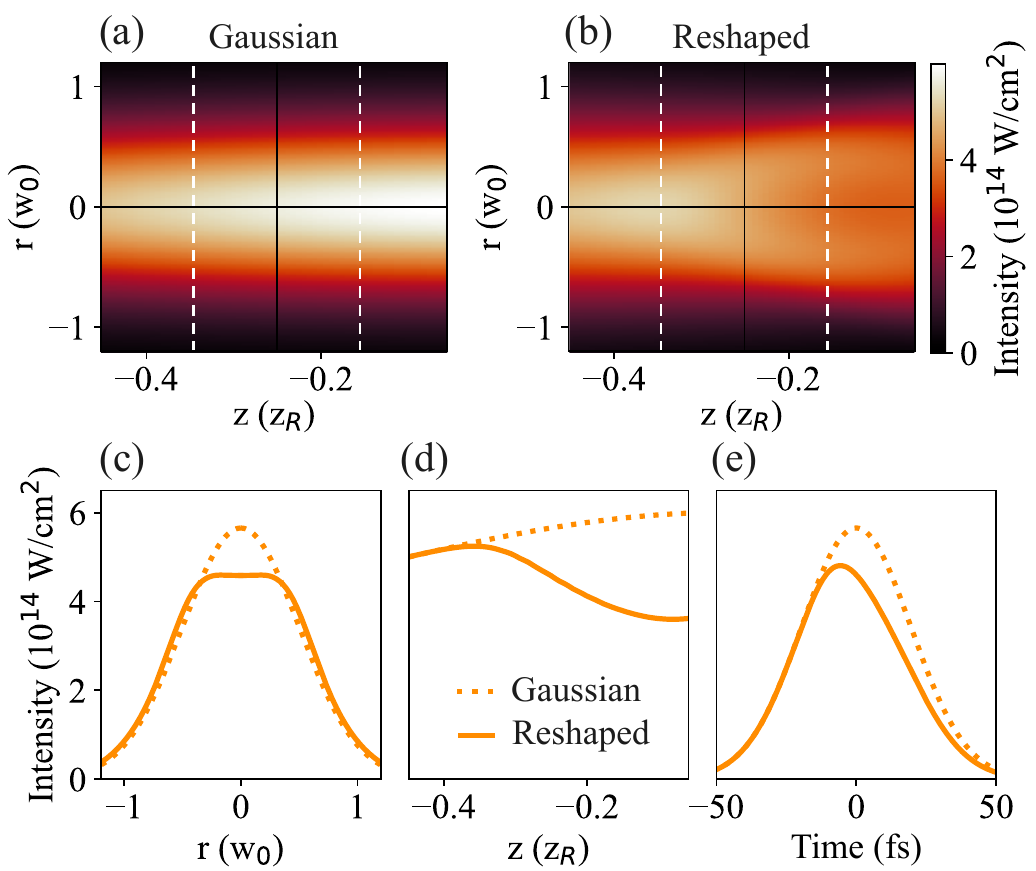}
    \caption{(a) Gaussian beam and (b) plasma-reshaped beam intensity at the peak of the pulse as it propagates through the medium. The white dashed lines represent the extent of the 600~$\mu$m gas jet. The black horizontal lines indicate $r=0$ and the vertical ones indicate the center of the gas jet at $z=-0.25z_{\mathrm{R}}$. $z=0$ is the geometrical focus of the Gaussian beam. (c) Radial profile of the Gaussian and reshaped beams at $z = -0.25z_{\mathrm{R}}$ and at $t = 0$. (d) Longitudinal profile of the Gaussian and reshaped beams at $r = 0$ and at $t = 0$. (e) Temporal profile of the Gaussian and reshaped pulses at $r = 0$ and $z = -0.25z_{\mathrm{R}}$.}
    \label{fig:reshaped_beam}
\end{figure}

The previous discussion assumes that the driving laser beam and pulse remain Gaussian throughout its propagation. In practice, the free electrons generated during HHG modify the refractive index of the medium, leading to a spatio-temporal reshaping of the fundamental field. As will be shown later, it is essential to include these effects in the phase-matching description in order to interpret our experimental results.  Figures \ref{fig:reshaped_beam}(a) and \ref{fig:reshaped_beam}(b) compare the propagation between a Gaussian beam and a plasma-reshaped beam close to the focal plane ($z=0$), obtained by numerical simulations described in Sec.~\ref{sec:simu}. For these calculations, we considered a 45~fs laser pulse with an energy of 550~$\mu$J and a central wavelength of 1032~nm, corresponding to the parameters used in our experiment. The longitudinal coordinate $z$ is expressed in units of the Rayleigh length $z_{\mathrm{R}}=3.2$~mm of the Gaussian beam, while the radial coordinate $r$ is in units of the beam waist $w_0=35$~$\mu$m in vacuum. The neon gas jet is 600~$\mu$m-long ($0.2 z_\mathrm{R}$) with a pressure of 500~mbar, placed 800~$\mu$m ($0.25 z_\mathrm{R}$) before the geometrical laser focus at $z=0$. Plasma reshaping results in a depletion of the beam on-axis [Fig.~\ref{fig:reshaped_beam}(c)], in a reduction of the intensity on-axis along $z$ [Fig.~\ref{fig:reshaped_beam}(d)], and a decrease of the intensity in time [Fig.~\ref{fig:reshaped_beam}(e)]. 

The reshaped beam can be analytically described using the formulation of refs.~\cite{akhmanovSelffocusing1968,aggarwalPropagationCoshGaussian2014}. The complex electric field is described as the product ${A=A_1(r,z)\exp[i\phi_1(r,z)]}$ with two real functions $A_1$ and $\phi_1$. Their evolution in a dilute medium is coupled in the paraxial equations:
\begin{align}
    &2\frac{\partial{\phi_1}}{\partial z}+\frac{1}{k_1}\left(\frac{\partial{\phi_1}}{\partial r} \right)^2 = \frac{1}{k_1A_1}\nabla^2_\perp A_1 - \beta\frac{r^2}{w_0^2},
    \label{eq:Aggarwal} \\
    &k_1\frac{\partial{A_1}}{\partial z} + \frac{\partial{A_1}}{\partial r}\frac{\partial{\phi_1}}{\partial r} =- \frac{A_1}{2} \nabla^2_\perp \phi_1,
\end{align}
where $\nabla^2_\perp$ is the Laplacian operator in the plane perpendicular to the propagation direction $z$ and $\beta$ is a parameter that describes the radially-dependent change of the refractive index $n$ due to plasma defocusing (${n^2 = n_0^2-\beta r^2/w_0^2}$). 
We consider Eq.~(\ref{eq:Aggarwal}) on-axis, which takes the simple form:
\begin{equation}
    \frac{\partial\phi_{\text{1}}}{\partial z} = \frac{1}{2k_1A_1}\nabla^2_\perp A_1\big|_{r=0}.
\label{eq:phi_1}
\end{equation}
In the special case of a Gaussian beam, Eq.~(\ref{eq:phi_1}) gives the derivative of the Gouy phase $-z_{\mathrm{R}}/(z^2+z_{\mathrm{R}}^2)$. This more general expression is essential to enable a description of phase-matching with a reshaped beam.

Plasma-induced reshaping does not occur instantaneously but develops as free electrons accumulate in the medium. The beam remains close to a Gaussian early in the pulse. As ionization increases, plasma defocusing progressively reshapes the beam, modifying both its wavefront curvature and its spatial intensity distribution. In the following, we use the calculated time-dependent intensity $I(r,z,t)=|A_1(r,z,t)|^2$ and the derivative of the on-axis phase obtained through Eq.~(\ref{eq:phi_1}) to extend the phase-matching description to include the effect of the plasma-induced reshaping of the fundamental beam.

\subsection{Phase-matching of harmonics with a plasma-reshaped field}

We now consider the effect of reshaping on the four terms in Eq.~(\ref{eq:delta k=0}). The contribution of the neutral atoms $\Delta k_{\mathrm{at}}$ does not change significantly in the presence of reshaping and remains time-independent, since the degree of ionization remains low.

\begin{figure}[htbp]
    \centering
    \includegraphics[width=0.99\linewidth]{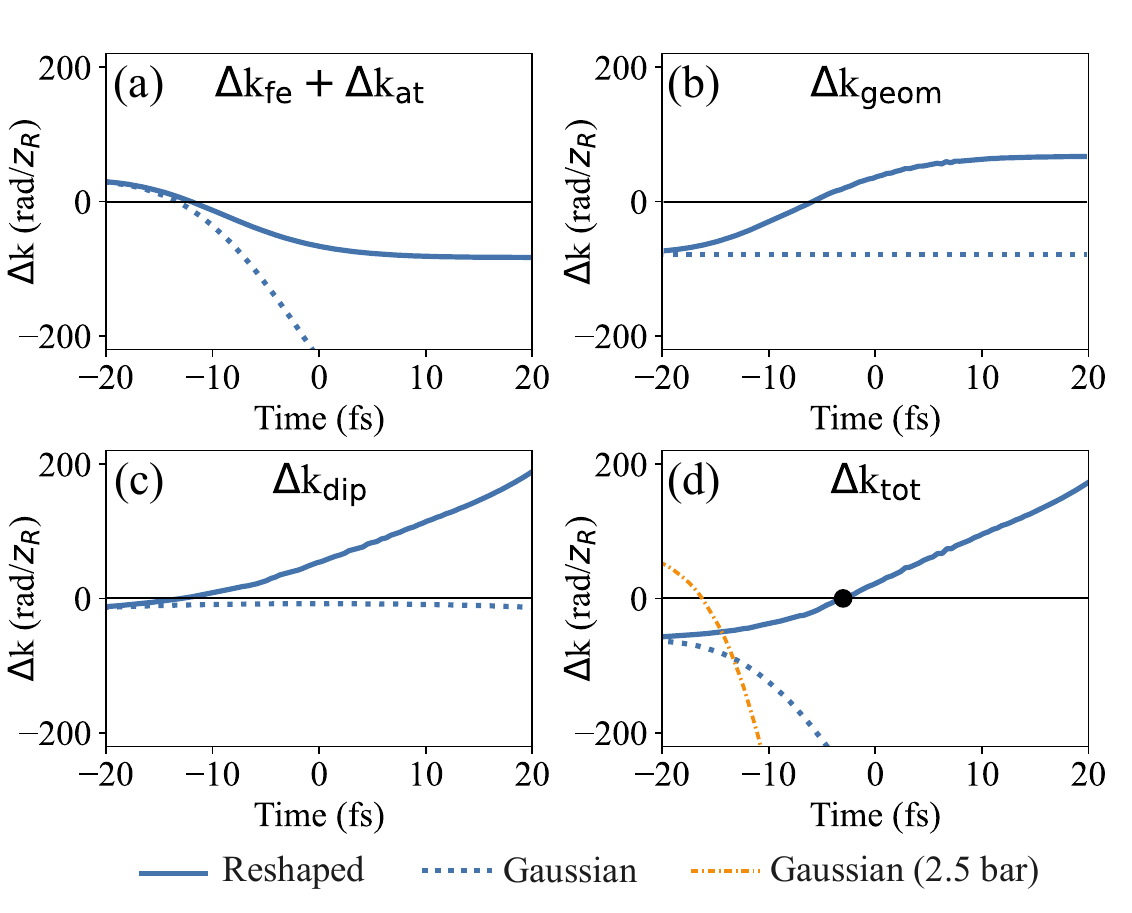}
    \caption{Comparison of the contributions to the phase-mismatch as a function of time, for (a) the free electrons and neutral atoms dispersion, (b) the geometrical phase-mismatch, (c) the short-trajectory dipole phase-mismatch and (d) the total mismatch. The pulse energy is 550~$\mu$J, the gas pressure is 500~mbar, and $z=-0.18z_{\mathrm{R}}$. The solid blue curves correspond to the plasma-reshaped beam, while the blue dotted curves show the Gaussian-beam case. The black dot highlights the time where phase-matching is achieved. The orange dot-dashed line in (d) represents the conventional phase-matching condition for a Gaussian beam, calculated for the same parameters but a higher gas jet pressure of 2.5~bar.}
    \label{fig:deltak_terms_as_function_of_time}
\end{figure}

Plasma defocusing lowers the intensity in time on-axis [Fig.~\ref{fig:reshaped_beam}(e)], resulting in a lower degree of ionization. Consequently, the time-dependent plasma-dispersion contribution $\Delta k_{\text{fe}}$ becomes less negative than for a Gaussian beam. Figure~\ref{fig:deltak_terms_as_function_of_time}(a) represents the variation of $\Delta k_{\text{fe}}(t)+\Delta k_{\text{at}}$ as a function of time.

The geometrical phase-mismatch term ${\Delta k_{\text{geom}}=q\partial \phi_1/\partial z}$ is represented in Fig.~\ref{fig:deltak_terms_as_function_of_time}(b). Compared to the Gaussian beam case, this term is modified through the factor $\nabla^2_\perp A_1$ in Eq.~(\ref{eq:phi_1}). $\Delta k_{\text{geom}}$ increases as a function of time, and can even become positive.

The dipole phase depends on both the laser intensity and the instantaneous driving frequency, so that $\Delta k_{\mathrm{dip}}$ can be written as:
\begin{equation}
    \Delta k_{\mathrm{dip}} = \frac{\partial \phi_{\mathrm{dip}}}{\partial I}\frac{\partial I}{\partial z} +\frac{\partial \phi_{\mathrm{dip}}}{\partial \omega_0}\frac{\partial \omega_0}{\partial z}.  
    \label{eq:deltak_dip}
\end{equation}
For few-cycle pulses and high ionization fractions, the blueshift of the driving field makes the second term dominant \cite{fuExtension2022,schotzPhaseMatching2020}. In our case, the longitudinal frequency gradient remains negligible compared to the intensity gradient. We therefore consider only the first term. Reshaping modifies the gradient $\partial I/\partial z$, leading to an increase of $\Delta k_{\mathrm{dip}}$ with time, that can become positive [Fig.~\ref{fig:deltak_terms_as_function_of_time}(c)].

For the conditions considered here, the total phase-mismatch is initially negative. Plasma-induced reshaping introduces positive contributions that grow over time and compensate for the negative mismatch, creating a time of efficient harmonic generation [black dot in Fig.~\ref{fig:deltak_terms_as_function_of_time}(d)]. Without considering reshaping, the contributions are time-independent except $\Delta k_{\mathrm{fe}}$, which decreases with time. Phase-matching can never be reached, as represented by the blue dotted line. To achieve phase-matching without reshaping requires a positive initial mismatch, which is possible for a different pressure or focusing geometry. This is exemplified by the yellow dotted line, obtained for the same geometry but with a pressure of 2.5~bar. The phase-matching window in this case occurs much earlier, on the rising edge of the pulse.

As shown in this discussion, the phase-matching condition selects a spatio-temporal window in which harmonic generation is efficient. Through the time-to-frequency mapping introduced above, this temporal window is directly encoded in the harmonic spectrum. Measuring harmonic spectral shifts provides experimental access to the temporal evolution of the phase-matching window, and therefore allows us to distinguish between the different phase-matching regimes.

\section{\label{sec:Methods}Methods and results}

\subsection{Experimental setup}

\begin{figure*}[htbp]
    \centering
    \includegraphics[width=0.95\linewidth]{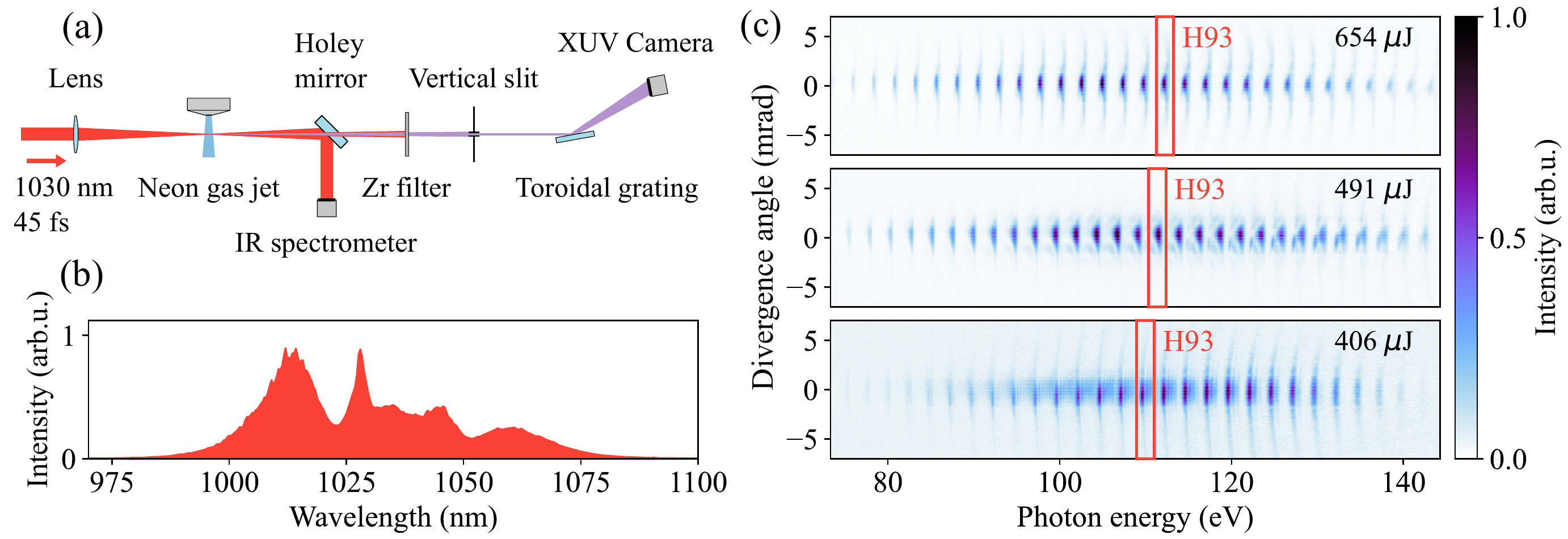}
    \caption{(a) Sketch of the experimental setup. An ultrashort laser from a post-compressed ytterbium source is focused by a convex lens into a neon gas jet. A holey mirror is used to filter most of the IR beam. The harmonics are then dispersed by a toroidal grating and imaged onto an XUV camera. Zr: Zirconium. (b) Example of an IR spectrum measured after the interaction with the gas jet. (c) Normalized harmonic spectra for three different driving pulse energies (406, 491 and 654~$\mu$J), at a constant gas jet pressure of 500~mbar. The XUV signal at harmonic 93 is lined out in red.}
    \label{fig:exp_setup}
\end{figure*}

The experimental setup is sketched in Fig.~\ref{fig:exp_setup}(a). High-order harmonics are generated in a neon gas jet using a post-compressed ytterbium laser. The laser delivers 45~fs pulses centered at 1032~nm with pulse energies up to 640~µJ. The pulse energy is tuned with an attenuator consisting of a motorized half-wave plate and a thin film polarizer.

The beam is focused with a convex lens to a focal spot of 35~$\mu$m in radius, located 800~$\mu$m ($0.25z_{\mathrm{R}}$) downstream of the neon gas jet that is used to produce the harmonics. The gas jet is produced by a 300~$\mu$m diameter round nozzle, and the backing pressure is varied up to 18~bar, corresponding to an estimated interaction length of 600~$\mu$m ($0.2z_{\mathrm{R}}$) and a jet pressure up to 500~mbar based on gas flow simulations.

After the interaction region, the residual infrared beam is filtered out with the help of a holey mirror and a zirconium filter. An infrared (IR) spectrometer is used at the reflection of the holey mirror to measure the spectrum of the residual IR beam [Fig.~\ref{fig:exp_setup}(b)]. The XUV beam is sent through a 50~$\mu$m vertical slit, and spectrally dispersed by a reflective toroidal grating. The slit is imaged onto an XUV camera, enabling the measurement of harmonics 59 to 119 with high spatial and spectral resolution. Spectra are recorded as a function of the pulse energy from 400 to 650~$\mu$J and the gas jet pressure from approximately 125 to 500~mbar.

\subsection{\label{sec:simu}Simulations}

In order to interpret the experimental observations, we perform numerical simulations of HHG in the neon gas jet. The single‑atom response is obtained from a pre‑computed table of ionization rates and dipole amplitudes and phases, calculated by solving the time‑dependent Schrödinger equation (TDSE) for a range of laser intensities and wavelengths. These TDSE data are then used as the microscopic source term in a macroscopic propagation model. The raw TDSE data contain phase components associated with longer electron trajectories that change by more than $2\pi$ with the intensity variation between consecutive half-cycles. To avoid associated numerical artifacts, the dipole response is filtered so that only the short and first long trajectories are retained.

The IR and harmonic fields are propagated in cylindrical symmetry under the slowly varying envelope approximation. The propagation of the IR accounts for the dispersion, diffraction, ionization, and Kerr effect. The propagation of the harmonic field only accounts for dispersion and diffraction. To compare directly with the measurements, the harmonic field is propagated to the far field using a Fourier transform of the field at the exit of the medium.

\subsection{Experimental and simulated results}

\begin{figure*}[htbp]
    \centering
    \includegraphics[width=0.99\linewidth]{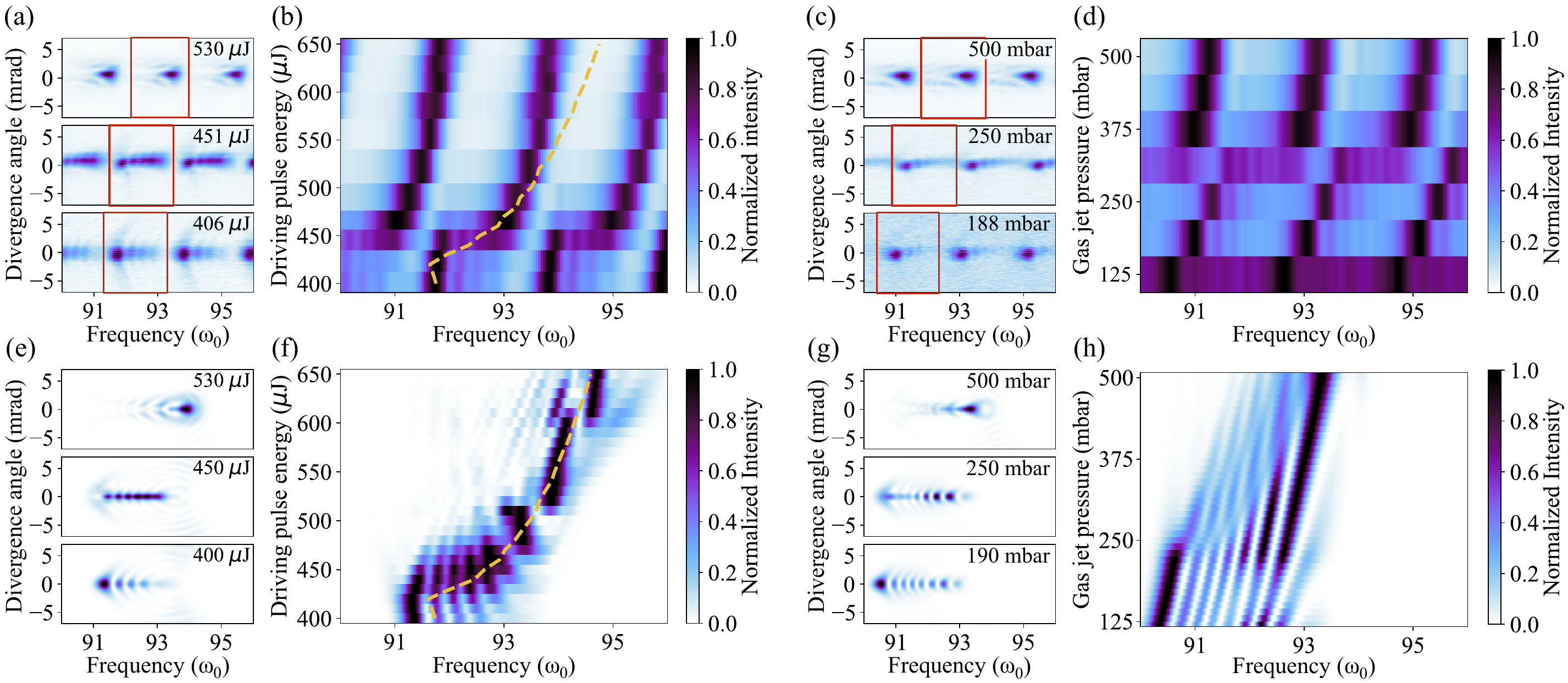}
    \caption{(a) Measured and (e) simulated far-field angularly resolved spectra for three selected pulse energies. \textbf{b} Measured and \textbf{f} simulated on-axis harmonic spectra as a function of the driving laser pulse energy. The yellow dashed line is the modeled instantaneous frequency of the harmonic at the moment of efficient emission. In (a), (b), (e) and (f), the gas pressure is 500~mbar. (c) Measured and (g) simulated far-field angularly resolved spectra for three selected gas jet pressures. (d) Measured and (h) simulated on-axis harmonic spectra as a function of the gas jet pressure. In (c), (d), (g) and (h), the pulse energy is 490 $\mu$J. Harmonic 93 is lined out in the red boxes in (c), (d) and it is the only harmonic considered in the simulations. $\omega_0$ is the fundamental frequency.}
    \label{fig:spectransition}
\end{figure*}

Figure~\ref{fig:exp_setup}(c) shows three spectra obtained at different driving pulse energies. The harmonic spatial profiles exhibit an intense central part and an extended outer off-axis region. Depending on the driving pulse energy, the two regions are centered at slightly different frequencies and with different spectral widths. In addition, the frequency of the harmonics strongly varies, by more than two harmonic orders, outlined by the red rectangle for harmonic 93.

In the following, we focus on harmonic 93. At a fixed gas jet pressure, we first investigate how the harmonic profile and spectrum evolve with increasing driving pulse energy. Figure~\ref{fig:spectransition}(a) highlights the profile of harmonic 93 at three different driving pulse energies and a pressure of 500~mbar. These profiles are also reproduced in the simulations using the TDSE pre-calculated data [Fig.~\ref{fig:spectransition}(e)]. The spatio-spectral profiles exhibit a series of interference circles, approximately centered around the same frequency. The most intense region is always on-axis and shifts to higher frequency as the energy increases, with a broad spectrum at 450~$\mu$J. Figures~\ref{fig:spectransition}(b) and \ref{fig:spectransition}(f) show respectively the measured and simulated on-axis harmonic spectra as a function of pulse energy. A pronounced spectral transition towards higher frequencies is observed around 450~$\mu$J. 

Figures~\ref{fig:spectransition}(c), \ref{fig:spectransition}(d), \ref{fig:spectransition}(g) and \ref{fig:spectransition}(h) show a similar behavior for a fixed pulse energy of 490~$\mu$J, for different gas pressures: the harmonic profiles also consist of interference rings, with an intense region shifting to the blue with increasing pressure. The on-axis spectrum presents an abrupt transition at a gas pressure of 300~mbar.

At the transition around 450~$\mu$J in Figs.~\ref{fig:spectransition}(b) and \ref{fig:spectransition}(f) and around 300~mbar in Figs.~\ref{fig:spectransition}(d) and \ref{fig:spectransition}(h), the spectrum is broad and exhibits pronounced interference fringes. Because of the strong chirp that gives rise to the time-to-frequency mapping mentioned previously, this bandwidth indicates in this case a harmonic emission over an extended temporal window, contrary to the usual Fourier-limited picture. The spectrum is narrower both below and above the transition, corresponding to a more confined harmonic temporal profile.

\section{\label{sec:Interpretation}Interpretation}

\subsection{Phase-matching-induced harmonic emission}

\begin{figure}[htbp]
    \centering
    \includegraphics[width=0.99\linewidth]{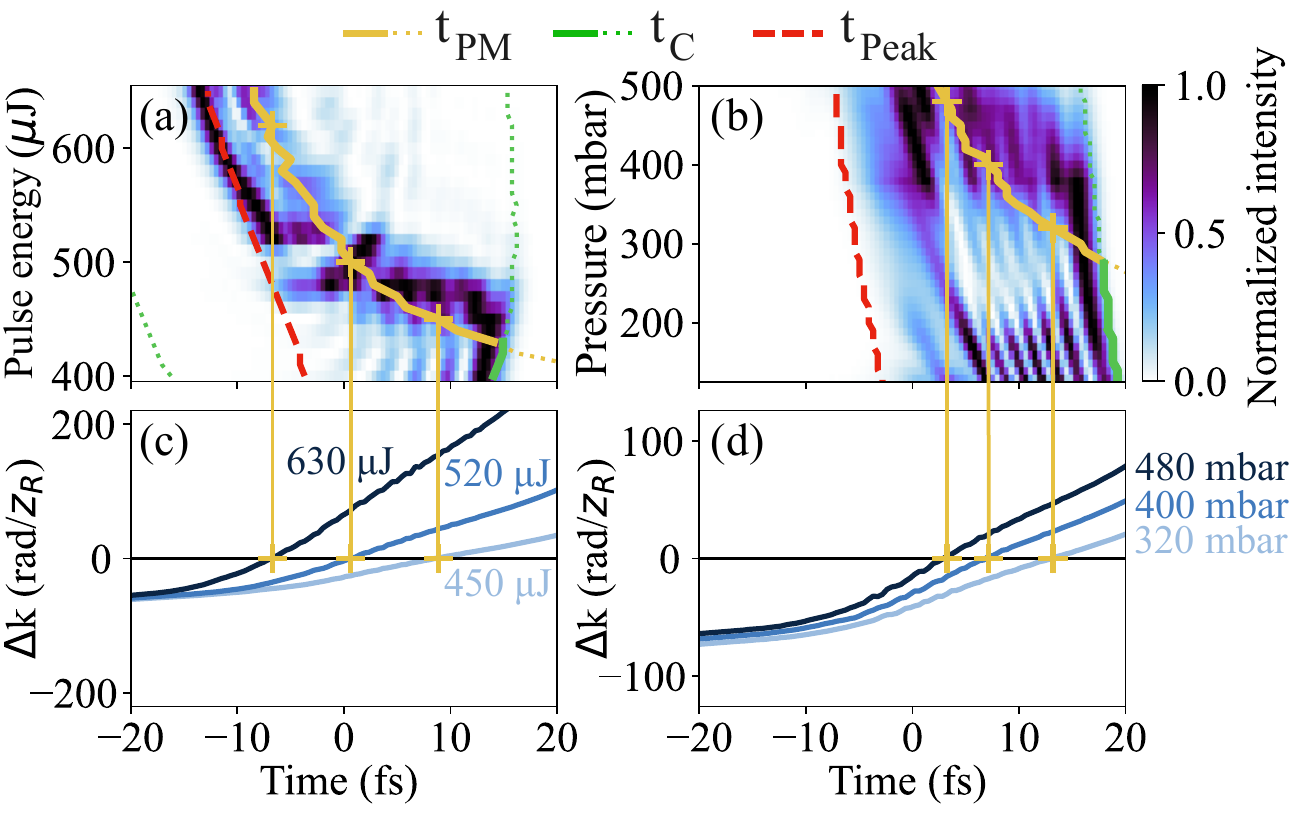}
    \caption{Simulated on-axis time profile of harmonic 93 as a function of (a) driving pulse energy and (b) gas pressure. The time $t_{\text{PM}}$ at which the total phase-mismatch cancels out for each energy is plotted in yellow, the times $t_{\mathrm{C}}$ at which the cut-off intensity is achieved is plotted in green, and the peak of the reshaped driving pulse $t_{\mathrm{Peak}}$ is in red. Total phase-mismatch as a function of time at $z=-0.18z_{\mathrm{R}}$, (c) for three different driving pulse energies at a constant gas pressure, and (d) for three different pressures at a constant pulse energy. The gas pressure is 500~mbar in (a) and (c), and the driving pulse energy is 490~$\mu$J in (b) and (d).}
    \label{fig:deltak_time}
\end{figure}

To understand the observed results, we study the harmonic emission as a function of time. Figures~\ref{fig:deltak_time}(a) and \ref{fig:deltak_time}(b) represent the simulated harmonic 93 temporal profiles as a function of the driving pulse energy for a gas pressure of 500~mbar (a) and as a function of the gas pressure for a driving pulse energy of 490~$\mu$J (b). Figure~\ref{fig:deltak_time}(a) is similar to Fig.~\ref{fig:spectransition}(f) with a flipped horizontal axis. This is a consequence of the time-to-frequency mapping mentioned earlier, in which a late emission corresponds to a spectrum shifted to the red relative to the nominal harmonic frequency (Fig.~\ref{fig:tf_mapping}). The observed intensity modulations are the result of quantum path interferences between the short and first long electron trajectories \cite{zairQuantumPathInterferences2008, augusteTheoreticalExperimentalAnalysis2009}, which are then mapped in the spectral domain in Fig.~\ref{fig:spectransition}. Figure~\ref{fig:deltak_time}(b) shows the same features as Fig.~\ref{fig:spectransition}(h) with a flipped horizontal axis. In this case, we also observe contributions from the long trajectory at pressures below 300~mbar. This is expected since the dipole phase mismatch of the long trajectory is more important than that of the short trajectory, enabling phase-matching at a lower pressure. The contributions from the two trajectories make it more difficult to apply the time-to-frequency mapping in this case. In the following, we focus on the short trajectory contribution, visible at high pressure.

The total phase-mismatch at a given position inside the jet is represented in Fig.~\ref{fig:deltak_time} as a function of time for different driving pulse energies at a gas pressure of 500~mbar (c) and for different gas pressures at a driving pulse energy of 490~$\mu$J (d). This phase-mismatch also depends on the longitudinal coordinate $z$, but for simplicity, we only consider the position $z=-0.18z_{\mathrm{R}}$, which corresponds to the region where the harmonic is mostly emitted. The moment at which the phase-mismatch cancels out defines a phase-matching time $t_{\mathrm{PM}}$. This time shifts to earlier times with increasing pulse energy and pressure because the effect of reshaping increases. As shown by the yellow line in Figs.~\ref{fig:deltak_time}(a) and \ref{fig:deltak_time}(b), $t_{\mathrm{PM}}$ follows the peak of the simulated normalized temporal profiles of harmonic 93 except at low driving pulse energy or low gas pressure. In these cases, phase-matching is achieved at late times, corresponding to a driving intensity that is already below the cutoff intensity required to generate harmonic 93. The time corresponding to the cutoff, $t_{\mathrm{C}}$, is indicated by the green lines in Figs.~\ref{fig:deltak_time}(a) and \ref{fig:deltak_time}(b). The time of efficient HHG, shown by the solid line, is then limited by the single-atom response rather than by phase-matching. Once phase-matching becomes possible above the cutoff intensity, the temporal window is no longer constrained by the single-atom cutoff and instead follows the phase-matching window.

\subsection{Harmonic and fundamental spectral shifts}

From the efficient emission times obtained [solid lines in Figs.~\ref{fig:deltak_time}(a)], we can calculate the corresponding harmonic frequency by using the time-to-frequency mapping described in Eq.~(\ref{eq:Om_inst}), with $r=0$ and $z=-0.18 z_{\mathrm{R}}$. This instantaneous frequency, shown as a dashed yellow line in Figs.~\ref{fig:spectransition}(b) and \ref{fig:spectransition}(f), reproduces the variation of the peak frequency with driving pulse energy in both simulation and experiment.

\begin{figure}[htbp]
    \centering
    \includegraphics[width=0.99\linewidth]{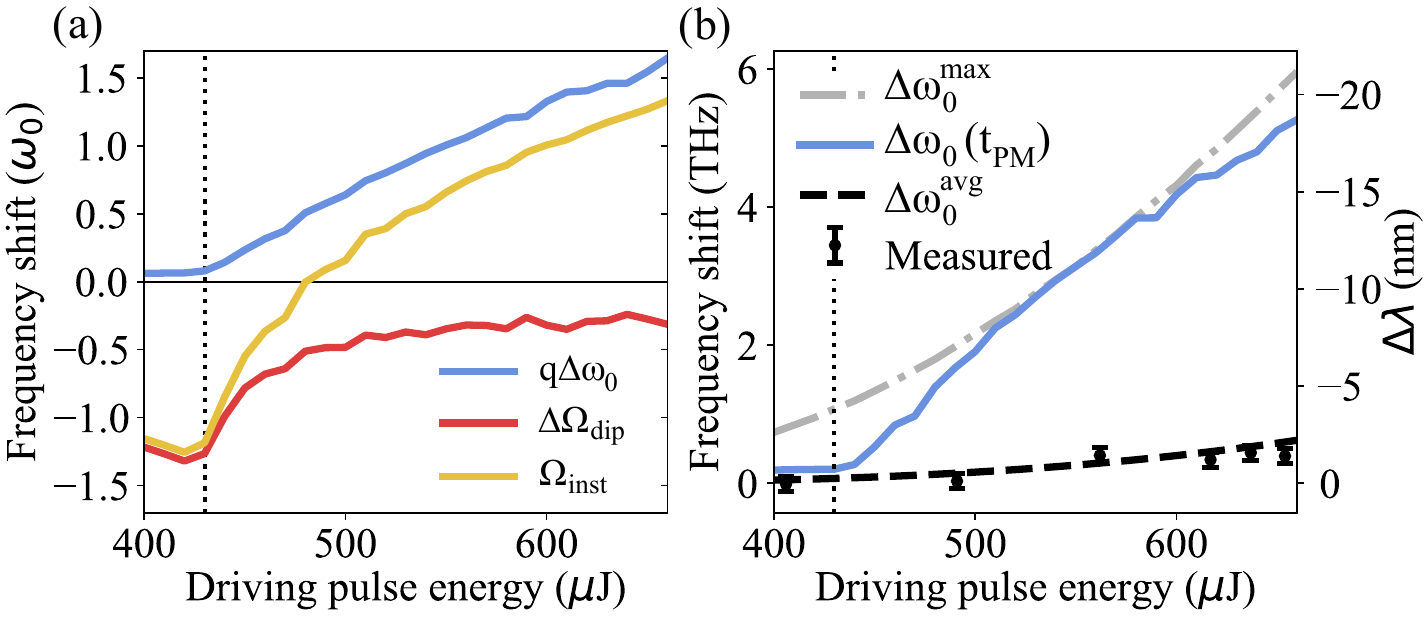}
    \caption{(a) Contributions to the spectral shift of harmonic 93 [Eq.~(\ref{eq:Om_inst})] at the time of efficient generation as a function of the driving pulse energy, for a gas pressure of 500~mbar. (b) Frequency shift of the driving laser field due to ionization for different pulse energies. The gray line represents the simulated maximum shift of the instantaneous frequency on axis $\Delta\omega_0^{\mathrm{max}}$, while the solid blue line represents the simulated instantaneous frequency shift at the moment of phase-matching. The black dashed line corresponds to the simulated time- and space-averaged spectral shift $\Delta\omega_0^{\mathrm{avg}}$ after the medium, which corresponds to what can be experimentally measured (markers). The vertical dotted lines show the energy at which $t_{\mathrm{PM}}=t_{\mathrm{C}}$, corresponding to the beginning of the transition.}
    \label{fig:contributions_spectral_shift}
\end{figure}

Figure \ref{fig:contributions_spectral_shift}(a) shows the relative contributions of the femto-chirp (red curve) and of the fundamental blueshift (blue curve) to the harmonic spectral shift (yellow curve) as a function of the driving pulse energy. In our conditions, the femto-chirp always leads to a redshift. This is a consequence of phase-matching being achieved after the peak of the reshaped driving pulse, which is shown by the red dashed line in Figs.~\ref{fig:deltak_time}(a) and \ref{fig:deltak_time}(b). The effect of the femto-chirp decreases from 430~$\mu$J to 500~$\mu$J as $t_{\mathrm{PM}}$ gets closer to the peak of the driving pulse [Fig.~\ref{fig:tf_mapping}]. While the redshift of the harmonic frequency is due to the femto-chirp, the observed blueshift at energies larger than 500~$\mu$J can be attributed to the shift of the fundamental (blue line). The observed spectral variation therefore results from the combined evolution of these two mechanisms. 

These results allow us to understand the origin of the spectral transition observed in Fig.~\ref{fig:spectransition}. When phase-matching becomes possible, $t_{\mathrm{PM}}$ shifts rapidly from late to early times, near the peak of the pulse. As temporal reshaping becomes more important, the driving pulse peak $t_{\mathrm{Peak}}$ moves to earlier times [Figs.~\ref{fig:deltak_time}(a) and \ref{fig:deltak_time}(b)], reducing the effect of the femto-chirp. Together with the increased contribution from the blueshift of the fundamental, a spectral transition is observed. The corresponding temporal window is large due to a slow variation of $\Delta k$ as a function of time [light blue curve in Fig.~\ref{fig:deltak_time}(c)]. Because of the time-to-frequency mapping discussed above, this extended efficient temporal window leads to a broad spectrum.

Finally, we discuss the amplitude of the blueshift of the fundamental in Fig.~\ref{fig:contributions_spectral_shift}(b). The dashed black line represents the simulated time- and space-averaged spectral shift of the driving pulse, $\Delta \omega_0^{\textrm{avg}}$, with the measured shift of the spectrum's center of mass shown as dots [Fig.~\ref{fig:exp_setup}(b)]. The maximum instantaneous spectral shift $\Delta \omega_0^{\textrm{max}}$ represented by the gray dot-dashed line is one order of magnitude stronger. The instantaneous frequency of the driving pulse at the efficient emission time of the harmonic, $\Delta \omega_{\mathrm{0}}(t_{\mathrm{PM}})$, is represented by the solid blue line. This comparison shows that the experimental spectrum cannot be used to estimate the blueshift that is transferred to the harmonics. What matters is the instantaneous frequency of the driving pulse at the instant of phase-matching, which is one order of magnitude stronger in this case. The gray dot-dashed curve and the blue line match for pulse energies above 500~$\mu$J, showing that the phase-matching time corresponds to the time at which the blueshift of the fundamental is maximum.

\subsection{Plasma-induced phase-matching regime}

\begin{figure}[htbp]
    \centering
    \includegraphics[width=0.9\linewidth]{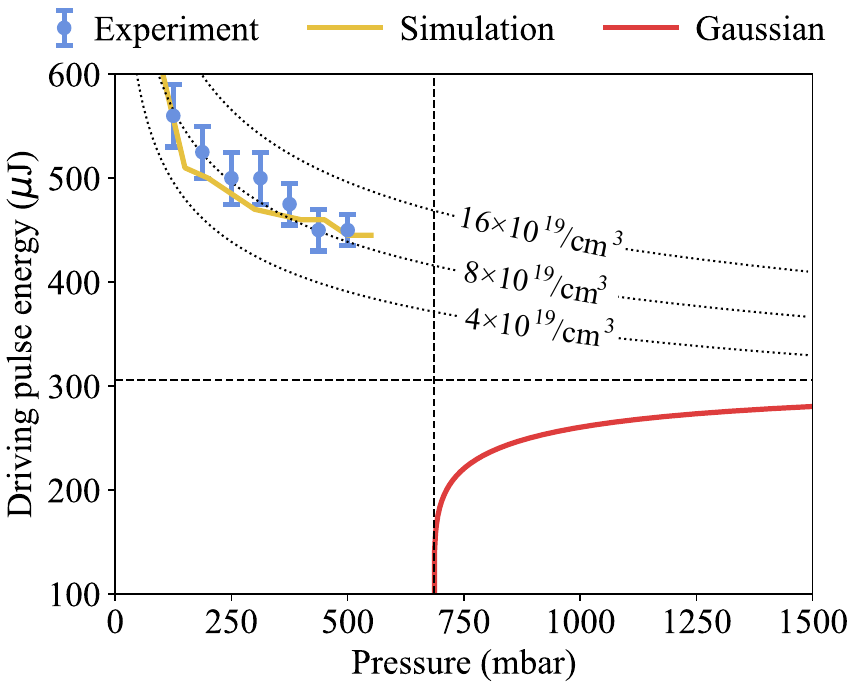}
    \caption{Measured (blue markers) and simulated (yellow line) points of observed spectral transition. The dotted curves represent curves of same electron density at the center of the gas jet and at $t=0$ for a Gaussian beam, in units of electrons/cm$^3$. Pulse energy required to achieve phase-matching at the peak of the Gaussian pulse without considering reshaping as a function of pressure (red curve). The horizontal black dashed line shows the energy above which phase-matching is not possible at the peak of the pulse in the Gaussian picture, while the vertical one shows the pressure under which phase-matching is not possible in the Gaussian picture.}
    \label{fig:E_PM_asfunctionof_p0}
\end{figure}

We recorded harmonic spectra as a function of the driving pulse energy for several gas pressures. For some pressures, we identified the spectral transition, where the spectrum is the broadest. These transition points are shown as blue markers in Fig.~\ref{fig:E_PM_asfunctionof_p0}, together with values obtained from the propagation simulations (yellow line). Since the spectral transition corresponds to the reshaping-dominated phase-matching regime discussed above, the transition points can be considered as points with similar phase-matching conditions. Because the amount of plasma-induced reshaping is determined by the free-electron density, these transition points are therefore expected to occur for similar electron densities.

We can estimate the pulse energy required to reach a given electron density as the gas pressure changes, following the derivation of \cite{minnekerCriticalLaserIntensity2023} (See Appendix~\ref{appendix:Minneker} for more details). The constant-electron-density curves plotted in dotted lines in Fig.~\ref{fig:E_PM_asfunctionof_p0} therefore correspond to curves of similar reshaping. The measured and simulated transition points closely follow the curve corresponding to an electron density of $8\times10^{19}$/cm$^{3}$, confirming that the phase-matching dynamics are primarily determined by the reshaping of the fundamental beam. 

This behavior is opposite to the conventional Gaussian phase-matching scenario \cite{gaardeMacroscopic2008,weissenbilderHowOptimizeHighorder2022} represented by the red curve in Fig.~\ref{fig:E_PM_asfunctionof_p0}, which is also obtained by following the derivation of \cite{minnekerCriticalLaserIntensity2023} in our geometry. In this case, a minimum pressure (vertical dashed line) is required so that the positive phase-mismatch coming from the dispersion can compensate the negative geometrical phase-mismatch (See Section \ref{sec:PM_Gaussian}). Increasing the gas pressure increases the positive contribution of dispersion to the phase-mismatch. Maintaining phase-matching therefore requires a higher pulse energy to increase the negative plasma-dispersion contribution. Phase-matching at the peak of the pulse is limited by a critical intensity (horizontal dashed line), above which the plasma-dispersion mismatch is too negative and phase-matching is impossible \cite{popmintchevBright2012,weissenbilderHowOptimizeHighorder2022}. In contrast, in the reshaping-dominated regime identified here, increasing the pressure requires lowering the driving pulse energy to maintain the same phase-matching, resulting in opposite behavior. Phase-matching is achieved below the critical pressure and above the critical pulse energy for conventional Gaussian phase-matching.

\section{Conclusion}

We have shown that the spectral shifts of high-order harmonics $(\approx93)$ reflect the temporal evolution of the phase-matching window in HHG. Increasing either the driving pulse energy or the gas pressure shifts this window towards earlier times within the laser pulse, producing spectral shifts spanning more than two harmonic orders. By comparing with propagation simulations, we show that this behavior is governed by plasma-induced reshaping of the driving field. Remarkably, reshaping remains the dominant mechanism controlling the phase-matching dynamics under the moderate intensities, low pressure, and short interaction length investigated here. An on-axis analytical description of the different phase-matching terms incorporating these effects successfully reproduces the trends observed in both the propagation simulations and the experimental measurements.

Our results highlight the importance of plasma-induced reshaping for the phase-matching of ytterbium-driven HHG under conditions relevant to compact, high-repetition-rate sources. In particular, they show that the conventional picture of a driving field that remains close to a Gaussian beam and pulse can be insufficient even at moderate intensities and over short interaction lengths. The broad harmonic spectra generated under certain conditions may also be of interest for applications requiring broadband XUV radiation or spectral tunability. On a broader scale, plasma-induced reshaping plays a key role in phase-matching at high photon energies and in the generation of isolated attosecond pulses. The intuitive description of the phase-mismatch terms for a reshaped field provided here offers a means of optimizing such HHG sources across the multi-parameter landscape and of engineering the interaction beyond conventional Gaussian-field descriptions.

\begin{acknowledgments}
C.L.A., A.L., A.-L.V. and P.E.-J. acknowledge support from the Swedish Research Council (Grant Nos. 2021-04691, 2021-05992, 2022-03519, 2023-04603 and 2023-04684). A.L. and P.E.-J. acknowledge support from the Knut and Alice Wallenberg Foundation (Grant Nos. 2020-0111 and 2023-0207) and the Wallenberg Center for Quantum Technology (WACQT). A.-L.V. acknowledges support from the Swedish Foundation for Strategic Research (Grant No. FFL24-0144) and Åforsk through the Swedish Foundations’ Starting Grant (Grant RAVIOLI). G.B., V.P. and S.W. acknowledge the support of the Helmholtz-Lund International Graduate School (HELIOS, Project No. HIRS-0018).
\end{acknowledgments}

\section*{Data availability}

The data that support the findings of this study are available from the corresponding author upon reasonable request.

\appendix*

\section{\label{appendix:Minneker}Electron density calculations}

To estimate the pulse energy required to reach a given electron density at the peak of a Gaussian pulse, we follow the derivation of \cite{minnekerCriticalLaserIntensity2023}. The required peak intensity $I$ needed to reach a given ionization degree $\eta$ at the peak of a Gaussian pulse of duration $\tau$,  is
\begin{align}
    &I=\left[ \frac{g_0}{F_0}W_0\left(-\frac{1}{g_0}\left(\frac{D_0}{\tau}\right)^{1/g_0} \right) \right]^{-2},\\
    &D_0=2\sqrt{2\ln(2)}\frac{|\ln(1-\eta)|}{\kappa_0}
\end{align}
In this expression, $I$ and $\tau$ are expressed in atomic units. $W_0$ is the real branch of the Lambert function, and the constants $g_0=-0.4114$, $F_0=1.3303$ and $\kappa_0=6.4204$ are defined for neon. 

We consider the limit of low degree of ionization and long pulses, so that $|\ln(1-\eta)|\approx\eta$, and $W_0(x)\approx\ln(x)$. The ionization degree can be expressed as a function of pressure $p_0$ as $\eta=n_ek_BT/p_0$ for an ideal gas where $n_e$ is the electron density, $k_B$ the Boltzmann constant and ${T=296}$~K the temperature. The peak intensity required to achieve a given electron density at the peak of a Gaussian pulse can therefore be estimated as
\begin{equation}
    I(p_0)=F_0^2\ln\left(2\sqrt{2\ln(2)}\frac{n_ek_BT}{\kappa_0\tau p_0}|g_0|^{|g_0|}\right).
\end{equation}
We then find the pulse energy in S.I. units with ${E(p_0)=I(p_0)\pi w_0^2 \tau/(2\times0.94)}$, and 1 a.u. of intensity equal to $3.51\times10^{16}$~W/cm$^2$.

\bibliography{RedshiftPaper} 

\end{document}